# Nonlinear high-frequency magnetic response of magnetoferritin metacrystals governed by spin thermodynamics

K.V.Kavokin,

*Spin Optics Laboratory, Saint-Petersburg State University, 198504, St. Petersburg, Russia, and Sechenov Institute of Evolutionary Physiology and Biochemistry, Russian Academy of Sciences, 194223, St. Petersburg, Russia.*

A theory is developed of the time-dependent magnetization of the metacrystal composed of magnetoferritin macromolecules. Such superstructures, comprising up to several millions of superparamagnetic nanoparticles encapsulated in protein shells, can be created artificially using biochemical assembling technologies. They have been also shown to occur naturally in sensitive cells of the inner ear of birds, which suggests their possible involvement in the detection of the geomagnetic field for orientation and navigation of migratory animals. The dynamics of the magnetic system of the magnetoferritin metacrystal, comprising a very large number of magnetic moments coupled by long-range dipole forces, is exceedingly complex. In order to find the response of the metacrystal to high-frequency magnetic fields, we used a thermodynamic approach borrowed from the theory of nuclear spin systems of solids. The resulted theory yields the time-dependent superspin temperature and magnetization induced by oscillating magnetic fields of arbitrary strength. The predicted dependence of the high-frequency response on the static magnetic field can be used for experimental detection and characterization of magnetoferritin metacrystals in biological tissues.

*1.Introduction.*

With the development of molecular assembly technologies, it has become possible to produce superstructures that mimic properties of crystalline solids on a greatly enhanced spatial scale. In particular, biotechnological self-assembling of magnetoferritin has made it possible to create metacrystals of hundreds of micrometers in size [1-6], the constant of their face-centered cubic (fcc) lattice being about 17 nanometers. Each node of such a metacrystal is the magnetoferritin macromolecule, which contains a magnetic core formed by a magnetite ($Fe_3O_4$) nanoparticle of nearly spherical shape and 7-9 nm in diameter, encapsulated by a non-magnetic protein shell (apoferritin) with the outer diameter of 12 nm [7-9] (see Fig.1). Magnetite particles of this size and shape are single-domain, demonstrating superparamagnetism down to blocking temperatures of about 20K [2]. As the magnetite cores are isolated from each other by apoferritin shells, coupling of their magnetic moments is possible only via the dipole-dipole magnetic interaction. Magnetoferritin metacrystals are considered prospective for engineering magnonic bandgap structures [3-4], as well as for biomedical applications [6]. They were found to occur naturally in mechanosensitive cells of the inner ear of birds [10-11], stirring the discussion about their possible role in magnetoreception [12]. The only qualitative analog of these superstructures in classical solid state physics is the nuclear spin system of a dielectric crystal, which also has a purely dipolar coupling. However, quantitative differences between the parameters of the two systems are huge: 7 orders of magnitude in the particle magnetic moment, about 1.5 orders in lattice constant, and over 9 orders in the spin-lattice relaxation time. The question therefore arises: to what extend the behavior of the nuclear spin system of a solid can be modelled with magnetoferritin metacrystals, and, vice versa, what can be adapted from the accumulated knowledge on the nuclear spin dynamics to get insight into this emerging class of artificial metamaterials?

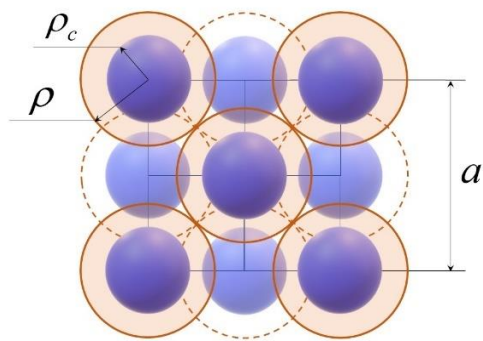

**Figure 1.** Structure of the magnetoferritin metacrystal. A fragment of the face centered cubic (fcc) lattice is shown in projection along one of the cubic axes. Blue spheres depict magnetite cores, apoferritin shells are shown by brown shading. $a \approx 17$ nm is the fcc lattice constant, $\rho \approx 6$ nm is the outer radius of the protein shell, and $\rho_c \approx 4$ nm is the radius of the magnetic core.

In this paper, we show that the magnetic response of the magnetoferritin metacrystal in the high-frequency (megahertz) domain is governed by the thermodynamics of the interacting system of the magnetic moments of constituent magnetite cores, thus bearing an analogy to the spin-temperature dynamics of the nuclear spin system of a solid. The collective dynamics results in a pronounced transient nonlinearity at room temperature, which can be used, in particular, for detection and characterization of natural or artificial magnetoferritin metacrystals in living organisms.

*2. Theory.*

The cores of magnetoferritin balls comprising the metacrystal are superparamagnetic particles with the saturation magnetization $M_S = 470$ G (in case of the pure magnetite monocrystal core), or somewhat less for cores of mixed composition (other iron oxides, ferrihydrite). For the core radius $\rho_c = 4$ nm it is easy to estimate that the magnetization saturation field at room temperature is about

1000 Gauss. The anisotropy energy of a spherical magnetic nanoparticle of this size is much less than the thermal energy $k_B T$ at room temperature. As it will be seen from the calculations presented below, the energy of magnetic interaction of a magnetic moment at a node of the metacrystal with magnetic moments at other nodes is also much less than $k_B T$. Therefore, the metacrystal remains in the paramagnetic state with random orientation of magnetic moments. These estimates are confirmed by measurements of static magnetization curves for different temperatures [2]: at 250K, the metacrystal magnetization is linear in magnetic field *B* up to *B*=500 G, with no remanence.

As distinct from the static susceptibility, the high-frequency susceptibility of a paramagnet may deviate from the Curie law even at high lattice temperatures. This was first noticed by Casimir and Du Pré back in 1938 [13]. The matter is that the exposure to varying magnetic field changes the energy of the paramagnet, pushing its magnetic system out of equilibrium with the crystal lattice. With increasing frequency, the isothermal susceptibility gives way to the adiabatic one, determined by the absence of energy transfer between the magnetic system and the lattice. The same should be true for the nonlinear susceptibility. In the following, a differential equation will be derived for magnetization induced by the external magnetic field that varies on the timescale longer than the time of establishing the internal equilibrium within the magnetic system. The equation will be verified by comparison with known results for the nuclear spin system of a classical solid, and then applied to describe the dynamics of magnetoferritin metacrystals.

The dynamics of the magnetic moment $\vec{m}_i$ of i-th node of the metacrystal, including the fluctuations of its direction, is described by the Langevin equation, based on the Landau-Lifshits-Gilbert equation [14]:

$$\frac{d\vec{m}_i}{dt} = -\gamma \left[ \vec{m}_i \times \left( \vec{B}_i + \vec{B}_{Ai} + \vec{b}_f \right) \right] - \gamma \eta \left[ \vec{m}_i \times \frac{d\vec{m}_i}{dt} \right] \qquad (1)$$

where $\vec{B}_i$ is the magnetic field at the node, $\vec{B}_{Ai} = \frac{\partial E_{Ai}}{\partial \vec{m}_i}$ is the anisotropy field, $E_{Ai}$ being the anisotropy energy of the magnetic core at this node, and $\vec{b}_f$ is the random field that takes into account the effect of lattice vibrations. The gyromagnetic ratio $\gamma$ is close to that of free electrons, as confirmed by experiments on microwave ferromagnetic resonance in magnetoferritin metacrystals [4]. The damping constant $\eta$ and the rms amplitude of the random field $\vec{b}_f$ for magnetic nanoparticles fixed in the metacrystal lattice are determined by coupling of the magnetic moment to vibrations [14]; the frequency spectrum of $\vec{b}_f$ at high temperatures is approximated by that of white noise. Solving Eq.(1) for an isolated magnetic nanoparticle yields the Néel relaxation time of its magnetic moment, $\tau_N$ [15]. In case of magnetic cores in the metacrystal, the magnetic field $\vec{B}_i = \vec{B} + \vec{b}_i$ includes, apart from the external magnetic field $\vec{B}$, also the field $\vec{b}_i$ created by other magnetoferritin particles via their dipole-dipole interaction. This fact greatly complicates the problem of the magnetic dynamics in the metacrystal as compared to one of an isolated magnetic nanoparticle, considered in Ref.[14], because $\vec{b}_i$ also fluctuates. Its dynamics follows fluctuations of magnetic moments of the other cores, which in turn are affected by fluctuating fields created by their neighbors. The interaction effects become the strongest when the Néel relaxation is slow, as expressed by the condition

$$\gamma B_{LK} \tau_N \gg 1, \qquad (2)$$

where $B_{LK} = \sqrt{\langle b_i^2 \rangle}$ is the kinetic local field, i.e. the mean squared field acting upon one of the core magnetic moments from all the other moments. In this limit, solving the dynamic problem becomes exceedingly difficult; however, as we show below, it can be approached thermodynamically. Indeed, if the condition given by Eq.(2) is fulfilled, the system of the magnetoferritin moments of the metacrystal becomes analogous to the nuclear spin system of a dielectric crystal. The nuclear spin system is known to reach the internal equilibrium within the time $T_2 \approx (\gamma_N B_{LN})^{-1}$, where $\gamma_N$ is the gyromagnetic ratio of the nuclear spin and $B_{LN}$ is the local magnetic field created by other nuclear spins. On the time scales longer than $T_2$, the nuclear spin system is described by the thermodynamic distribution function (or, quantum-mechanically, the density matrix) defined by the spin temperature $\Theta_N$, which may differ from the lattice temperature by orders of magnitude. The magnetization in a slowly (as compared to $T_2$) varying magnetic field at each moment of time is given by the Curie law with the instantaneous spin temperature. The dynamics of spin temperature is determined by energy redistribution between Zeeman and spin-spin interaction reservoirs as well as by the spin-lattice relaxation, i.e. levelling of the spin temperature with the temperature of the lattice [16-17]. To estimate applicability of this approach to the magnetoferritin metacrystal, let us calculate the kinetic local field $B_{LK}$, i.e. the mean squared field acting upon one of the core magnetic moments from all the other moments, assuming their random orientation:

$$B_{LK}^2 = \sum_j \left(\frac{1}{r_j}\right)^6 \left\langle \left( \vec{m}_j - 3\frac{\vec{r}_j(\vec{m}_j \cdot \vec{r}_j)}{\vec{r}_j^2} \right)^2 \right\rangle =$$

$$= \sum_j \left(\frac{1}{r_j}\right)^6 \left( m^2 - 6\frac{\left\langle (\vec{m}_j \cdot \vec{r}_j)^2 \right\rangle}{r_j^2} + 9\frac{\left\langle (\vec{m}_j \cdot \vec{r}_j)^2 \right\rangle}{r_j^2} \right) = \quad (3)$$

$$= \sum_j \left(\frac{1}{r_j}\right)^6 \left( m^2 + 3\frac{m^2 r_j^2 / 3}{r_j^2} \right) = 2m^2 \sum_j \left(\frac{1}{r_j}\right)^6$$

where the vector $\vec{r}_j$ define the position of j-th node with respect to the selected one. Numerical summation over the fcc lattice yields $\sum_j \left(\frac{1}{r_j}\right)^6 \approx 115.7 a^{-6} \approx 0.226 \rho^{-6}$, where $a = 2\sqrt{2}\rho$ is the fcc lattice constant, and $\rho$ is the external radius of the apoferritin shell. Expressing the magnetic moment via the saturation magnetization and volume of the magnetite core as $m = \frac{4\pi}{3}\rho_c^3 M_S$, we obtain

$$B_{LK} \approx 2.82 \frac{\rho_c^3}{\rho^3} M_S \quad (4)$$

Taking $\rho = 6nm$ and $\rho_c = 4nm$, we find $B_{LK} \approx 0.83 M_S$. Correspondingly, $T_2 \approx (\gamma B_{LK})^{-1} \approx 0.15 ns$ for $M_S = 470$ G.

The approach of the magnetic system to the internal equilibrium may be slowed down by anisotropy of magnetite cores, and this is indeed expected to happen at lower temperatures, close to the blocking temperature of about 20K [2]. In analogy to the Néel relaxation [14-15], the dynamics of the projection of the magnetic moment of a specific core on its anisotropy axis is slowed down by the factor $\exp(E_A / k_B T)$, where $E_A$ is the anisotropy energy. However, at room temperature this effect should be negligible, as the exponential factor is close to unity. It is worth noting that a similar problem exists in the nuclear spin physics, due to quadrupole interaction of nuclear spins with electric field gradients induced by strain, which results in the same dependence of energy on the direction of the particle magnetic moment as that in case of the uniaxial anisotropy of magnetic nanoparticles. As it was experimentally shown in semiconductor microstructures under optical cooling of the nuclear spin system [18], the dipole-dipole interaction is still able to establish the equilibrium within the spin system even when the effective quadrupole field (an analog of the anisotropy field here) is 5-8 times larger than the dipole-dipole local field.

This value of $T_2$ should be compared with the Néel relaxation time $\tau_N$, which is an analog of the spin-lattice relaxation time $T_1$ of the nuclear spin system. In case of larger superparamagnetic particles with large uniaxial anisotropy ($E_A/k_B T \gg 1$), the Néel relaxation time is given by an approximate formula $\tau_N \approx \tau_0 \sqrt{\pi k_B T / 2 E_A} \exp(E_A / k_B T)$ [14,19], where $\tau_0 \sim 1\,\text{ns}$ [20-21]. For the relevant case of low anisotropy and high temperature, no universal theoretical expression exists to the best of our knowledge, so that it is better to rely on experimental data. In Ref.[22], $\tau_N$ in dispersed magnetoferritin was determined by fitting the frequency dependence of the efficiency of its heating by oscillating magnetic fields [19]. For samples with $\rho_c = 4.3\,nm$ and $\rho_c = 4.8\,nm$, $\tau_N$ of 11 and 75 ns was found, respectively. This result suggests that the condition $\tau_N \gg T_2$ is likely to be fulfilled for magnetoferritin metacrystals at room temperature at least.

The other condition of the validity of the thermodynamic approach is that the external field $B$ should not change too fast, which is expressed by the inequality $2\pi f T_2 \ll 1$, where $f$ is the frequency of variation of $B$. From this condition, we find that the magnetic system of the metacrystal maintains the internal equilibrium if the frequency spectrum of the external perturbation does not extend above hundreds of megahertz.

Given all the necessary conditions for using the thermodynamic approach are satisfied, one can pass to deriving the equation which would describe the magnetization dynamics of the metacrystal in magnetic field of varying strength. The key point here is that the distribution function of the magnetic system is assumed to be determined by a single parameter, which, in analogy with the nuclear spin system, we will call superspin temperature, $\theta_{SS}$. In fact, it is more convenient to use the inverse superspin temperature determined as $\beta \equiv \dfrac{1}{k_B \theta_{SS}}$, where $k_B$ is the Boltzmann constant.

In the following, we assume that $k_B \theta_{SS}$ is much larger than the characteristic energy per one magnetite particle in the metacrystal.

The internal energy of the magnetic system is obtained by averaging the energy over configurations of the magnetic moments with the distribution function defined by $\beta$:

$$E = \sum_\sigma f_\sigma E_\sigma = \Xi^{-1} \sum_\sigma (1 - \beta E_\sigma) E_\sigma = -\beta \Xi^{-1} \sum_\sigma (E_Z + E_{dd} + E_A)^2 = -\beta \Xi^{-1} \sum_\sigma (E_Z^2 + E_{dd}^2 + E_A^2) \quad (5)$$

where $f_\sigma = \dfrac{\exp(-\beta E_\sigma)}{\sum_\sigma \exp(-\beta E_\sigma)} \approx \dfrac{1}{\Xi}(1-\beta E_\sigma)$ is the probability to occupy the state $\sigma$, $\Xi$ is the total number of states, and zero energy corresponds to the completely disordered magnetic system. Here the Zeeman energy is defined as

$$E_Z = \sum_i \vec{m}_i \cdot \vec{B} \qquad (6)$$

where $\vec{m}_i$ are vectors of magnetic moments of the cores. The last equality in Eq.(5) holds because cross-products of the Zeeman energy $E_Z$, the dipole-dipole energy $E_{dd}$ and the anisotropy energy $E_A$ vanish upon averaging over the states of the uncorrelated system of magnetic moments. Substituting Eq.(5) into Eq.(6), we obtain

$$E = E_Z + E_{dd} + E_A = -\beta \Xi^{-1}\left(\sum_\sigma \sum_{i,j}(\vec{m}_i \cdot \vec{B})(\vec{m}_j \cdot \vec{B}) + \sum_\sigma (E_{dd}^2 + E_A^2)\right) = -\beta N \cdot \frac{1}{3}m^2(B^2 + B_L^2) \qquad (7)$$

where $E_Z$, $E_{dd}$ and $E_A$ are mean values of Zeeman, dipole-dipole and anisotropy energies, $N$ is the total number of particles in the metacrystal, and the squared thermodynamic local field $B_L^2$ by definition equals

$$B_L^2 \equiv \frac{3}{Nm^2}\Xi^{-1}\sum_\sigma (E_{dd}^2 + E_A^2) \qquad (8)$$

Here

$$\Xi^{-1}\sum_\sigma E_{dd}^2 = \frac{1}{2}\left\langle \left[\sum_{i,j}\frac{1}{r_{ij}^3}\left(\vec{m}_i\vec{m}_j - 3(\vec{m}_i\cdot\vec{r}_{ij})(\vec{m}_j\cdot\vec{r}_{ij})/r_{ij}^2\right)\right]^2 \right\rangle =$$
$$= \frac{1}{2}\sum_{i,j}\left[\frac{1}{r_{ij}^3}\left\langle \vec{m}_i\cdot(\vec{m}_j - 3\vec{r}_{ij}(\vec{m}_j\cdot\vec{r}_{ij})/r_{ij}^2)\right\rangle\right]^2 = \frac{Nm^2}{6}B_{LK}^2 \qquad (9)$$

where angular brackets denote independent angular averaging of each $\vec{m}_i$, and $\Xi^{-1}\sum_\sigma E_A^2$ is obtained in analogous way by averaging squared anisotropy energies at all the nodes.

The mean value of the total magnetic moment of the metacrystal is

$$M_B = \sum_\sigma f_\sigma M_{B\sigma} = \Xi^{-1}\sum_\sigma (1-\beta E_\sigma)M_{B\sigma} = \beta B \Xi^{-1}\sum_\sigma M_{B\sigma}^2 = \frac{1}{3}\beta N m^2 B \qquad (10)$$

The rate of changing the energy of a closed system, $\dot{E}$ (in our case it is the rate of changing the energy of the metacrystal as a whole; the heat exchange with the environment can be neglected on the sub-microsecond time scale) under variation of external parameters is equal to the mean value of the partial time derivative of the system Hamiltonian [23]. In our particular case, since the only parameter explicitly depending on time is the external magnetic field $B$, this rate equals

$$\dot{E} = \left\langle \frac{\partial \hat{H}}{\partial B} \right\rangle \frac{dB}{dt} = \frac{\partial E_Z}{\partial B} \frac{dB}{dt} = -M_B \frac{dB}{dt} = -\frac{1}{3} \beta N m^2 B \frac{dB}{dt} \tag{11}$$

The rate of changing the total energy should be equal to the sum of the rate of changing the internal energy of the magnetic system given by Eq.(7) as a result of variations of the magnetic field and of the superspin temperature, and of the rate of energy exchange with the lattice:

$$\dot{E} = \frac{d}{dt}\left[-\beta N \cdot \frac{1}{3}m^2\left(B^2 + B_L^2\right)\right] - \frac{1}{\tau_N}\frac{\partial E}{\partial \beta}(\beta - \beta_L) =$$
$$= -\frac{Nm^2}{3}\left[\left(B_L^2 + B^2\right)\frac{d\beta}{dt} + 2\beta B \frac{dB}{dt}\right] + \frac{1}{\tau_N} \cdot \frac{Nm^2}{3}\left(B^2 + B_L^2\right)(\beta - \beta_L) \tag{12}$$

where $\beta_L \equiv (k_B T_L)^{-1}$ is the inverse lattice temperature.

The requirement that right-hand sides of Eqs. (11) и (12) should be equal to each other yields the differential equation for $\beta$:

$$\frac{d\beta}{dt} = -\beta \frac{B}{B^2 + B_L^2}\frac{dB}{dt} - \frac{\beta - \beta_L}{\tau_N} \tag{13}$$

Note that $\frac{B}{B^2 + B_L^2}\frac{dB}{dt} = \frac{1}{2}\frac{d}{dt}\ln\left(B_L^2 + B^2\right) = \frac{d}{dt}\ln\sqrt{B_L^2 + B^2}$. At $\tau_N \to \infty$ Eq.(13) reduces to the equation $\frac{d}{dt}\ln\beta = -\frac{d}{dt}\ln\sqrt{B_L^2 + B^2}$, by solving which one arrives to the well known formula for the adiabatic variation of the inverse spin temperature, usually obtained from entropy conservation in the adiabatic process [16,17]:

$$\beta(t) = \beta(0)\sqrt{\frac{B_L^2 + B^2(0)}{B_L^2 + B^2(t)}} \tag{14}$$

In order to find the deviation of the superspin temperature from the lattice temperature under the time-dependent magnetic field, we rewrite Eq.(13) the following way:

$$\frac{d\beta_m}{dt} = -\beta_m \left( \frac{1}{\tau_N} + \frac{d}{dt} \ln g(t) \right) - \beta_L \frac{d}{dt} \ln g(t) \qquad (15)$$

where $\beta_m = \beta - \beta_L$, and $g(t) = \sqrt{B_L^2 + B^2(t)}$.

The solution to Eq.(15) has the form

$$\beta_m(t) = \int_{-\infty}^{t} \left( -\beta_L \frac{d}{dt'} \ln g(t') \right) G(t,t') dt' \qquad (16)$$

where the Green function $G(t,t')$, found from the equation

$$\frac{dG(t,t')}{dt} = -G(t,t') \left( \frac{1}{\tau_N} + \frac{d}{dt} \ln g(t) \right) + \delta(t-t') \qquad (17)$$

equals

$$G(t,t') = \frac{g(t')}{g(t)} \exp\left( -\frac{t-t'}{\tau_N} \right) \qquad (18)$$

Substituting Eq.(18) into Eq.(16) yields

$$\begin{aligned}
\beta_m(t) &= -\beta_L \int_{-\infty}^{t} \left[ \frac{d}{dt'} \ln g(t') \frac{g(t')}{g(t)} \exp\left( -\frac{t-t'}{\tau_N} \right) \right] dt' = \\
&= -\frac{\beta_L \exp(-t/\tau_N)}{g(t)} \int_{-\infty}^{t} \frac{dg(t')}{dt'} \exp(t'/\tau_N) dt' = \\
&= -\frac{\beta_L \exp(-t/\tau_N)}{g(t)} \left[ g(t) \exp(t/\tau_N) - \frac{1}{T_1} \int_{-\infty}^{t} g(t') \exp(t'/\tau_N) dt' \right] = \\
&= -\beta_L + \frac{\beta_L}{\tau_N g(t)} \int_{0}^{\infty} g(t-\tau) \exp(-\tau/\tau_N) d\tau
\end{aligned} \qquad (19)$$

Finally, we obtain

$$\begin{aligned}
\beta(t) &= \frac{\beta_L}{\tau_N g(t)} \int_{0}^{\infty} g(t-\tau) \exp(-\tau/\tau_N) d\tau = \\
&= \frac{\beta_L}{\tau_N \sqrt{B_L^2 + B^2(t)}} \int_{0}^{\infty} \sqrt{B_L^2 + B^2(t-\tau)} \exp(-\tau/\tau_N) d\tau
\end{aligned} \qquad (20)$$

The magnetic moment is then found using Eq.(10):

$$M_B = \frac{Nm^2}{3}\beta(t)B(t) = \frac{Nm^2\beta_L B(t)}{3\tau_N \sqrt{B_L^2 + B^2(t)}} \int_0^\infty \sqrt{B_L^2 + B^2(t-\tau)}\exp(-\tau/\tau_N)d\tau \quad (21)$$

## 3. Numerical examples of the high-frequency magnetic response.

Eq.(21) allows one to calculate the response of the magnetic system of the metacrystal to oscillating magnetic fields of arbitrary strength, by means of numeric evaluation of the integral. In case of the periodic external field $B(t) = B(t-T_B)$, the time integral in Eqs.(20) and (21) is converted into an integral over the period of the field variation, $T_b$:

$$\int_0^\infty \sqrt{B_L^2 + B^2(t-\tau)}\exp(-\tau/\tau_N)d\tau = \int_0^{T_b}\sqrt{B_L^2 + B^2(t-\tau)}\exp(-\tau/\tau_N)d\tau \cdot \sum_{n=0}^\infty \exp(-nT_b/\tau_N) =$$
$$= \frac{1}{1-\exp(-T_b/\tau_N)}\int_0^{T_b}\sqrt{B_L^2 + B^2(t-\tau)}\exp(-\tau/\tau_N)d\tau \quad (22)$$

Having chosen $B(t)$ as $B(t) = B_0 + b\sin(\omega t)$, we find $\beta(t)$ and $M_B(t)$ (see examples in Fig.2).

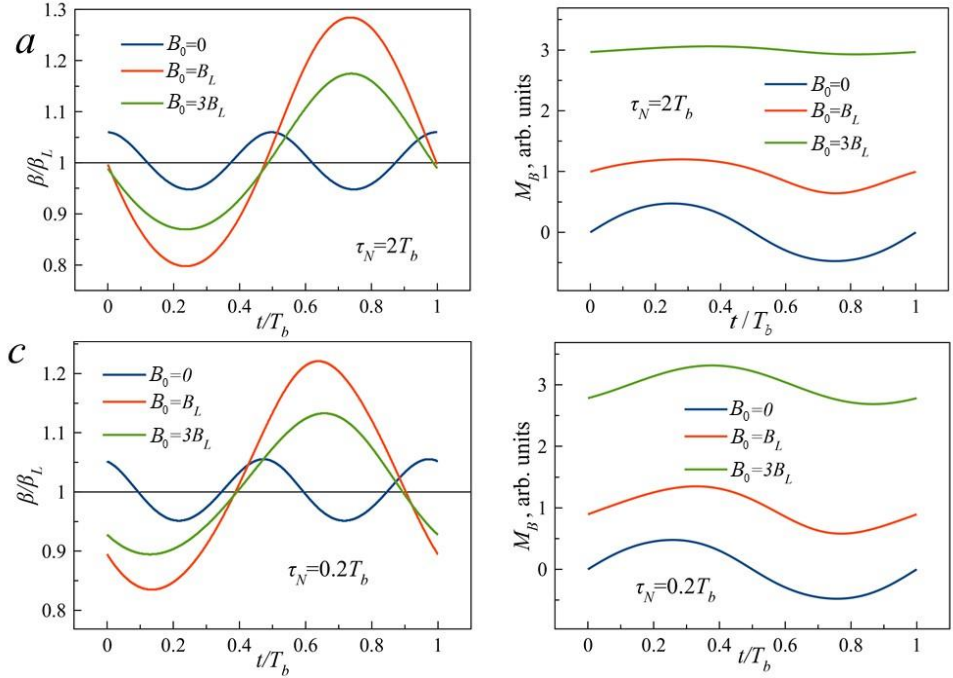

**Fig.2.** Examples of $\beta$ (*a* and *c*) and $M_B$ (*b* and *d*) as functions of time over the period of variation of the magnetic field $B(t) = B_0 + b\sin(\omega t)$, for $b = 0.5B_L$ and $B_0$ from 0 to $3B_L$ (as shown in panels), $\tau_N = 2T_b$ (*a* and *b*) and $\tau_N = 0.2T_b$ (*c* and *d*).

In case of slow spin-lattice relaxation ($\tau_N = 2T_b$), the dependence of $\beta$ on the magnetic field is close to the adiabatic one, given by Eq.(14). In this case, at $B \gg B_L$ $\beta$ is inversely proportional to

$B$, and the field dependence of magnetization becomes flat, which is seen from its weak response to the oscillating field at $B_0 = 3B_L$. The nonlinearity of the response is clearly seen, especially in the $M_B(t)$ curve for $B = B_L$. At fast spin-lattice relaxation ($\tau_N = 0.2T_b$) these effects become weaker, along with arising of a phase shift, better seen in $\beta(t)$ curves.

Let us analyze the non-linear effects in the response in more detail. Particularly, we will focus on the 2nd harmonic of $M_B(t)$, which can be detected experimentally with high sensitivity using a selective amplifier in order to suppress undesirable interference from the induction coil [24]. In particular, measuring the real and imaginary parts of the complex amplitude of the second harmonic defined as [25]:

$$\operatorname{Re} M_2(B) = -\frac{2}{T_b}\int_0^{T_b} M_B(t)\cos(2\omega t)dt$$
$$\operatorname{Im} M_2(B) = -\frac{2}{T_b}\int_0^{T_b} M_B(t)\sin(2\omega t)dt$$
(23)

as a function of the constant magnetic field $B$, has allowed the authors to realize a sensitive method for detection of magnetic nanoparticles dispersed in various media, e.g. in biological tissues [26, 27]. In Fig.3, examples of such curves are given for two different values of the ratio $\tau_N / T_B$.

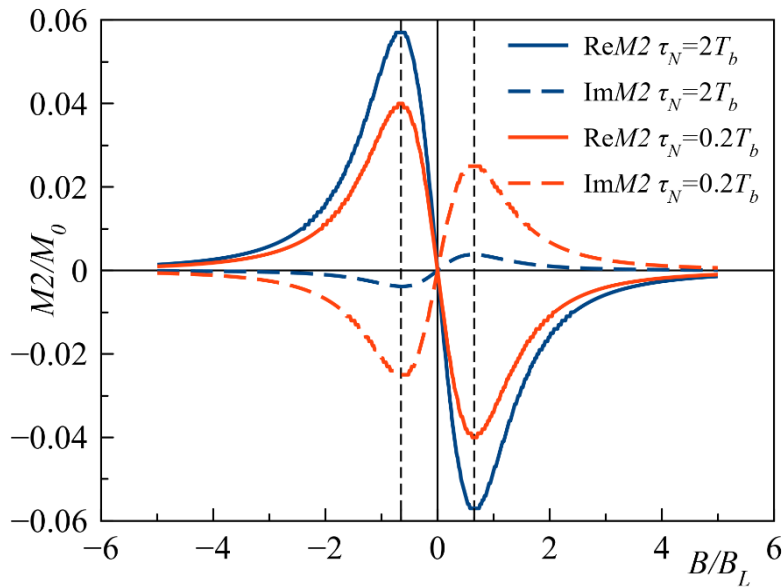

**Fig.3**. Real and imaginary parts of the second harmonic of nonlinear magnetic response of the magnetoferritin metacrystal, in the units of $M_0 = \frac{\beta_L B_L N m^2}{3} = \frac{B_L N m^2}{3k_B T}$. $b = 0.5 B_L$, $\tau_N = 2T_B$ and $0.2T_B$. Vertical dashed lines show the positions of extrema of the curves at $B_0 / B_L \approx \pm 0.65$.

The second harmonic amplitude reaches its maximum at $|B_0| \approx 0.65 B_L$. The amplitude of the imaginary part strongly increases with shortening of $\tau_N$, reflecting the growing phase shift of the nonlinear response. These features can be used for experimental determination of $B_L$ and $\tau_N$. In turn, the local field $B_L$ gives an estimate of the interaction strength in the metacrystal via Eq.(8), though in order to distinguish between the dipole-dipole interaction and the anisotropy of magnetoferritin cores, additional information is needed.

*4. Discussion.*

The developed theory suggests several directions of experimental research.

Experiments on the response of the metacrystal to high-frequency magnetic fields at room temperature or higher temperatures, compared with the above theory, can be helpful for characterization of metacrystals, in particular for determination of the magnetic moment and anisotropy energy of magnetoferritin cores. Even more important is experimental verification of the thermodynamic approach to the dynamics of the system of macroscopic magnetic moments. It is so far well established for quantum spins of nuclei; even in that case, Eq.(14) for the adiabatic variation of spin temperature with slowly changing magnetic field had not been verified in the direct experiment until recently [18]. A similar experimental study for the spatially ordered system of classically large magnetic moments would be a noticeable contribution to the thermal physics in general.

Experiments at lower temperatures may appear very interesting for the physics of magnetic phase transitions. The nuclear spin system of a solid is known to demonstrate phase transitions into magnetically ordered phases when its spin temperature is lowered down to the nanokelvin range [28, 29], which is a major technical challenge. With the magetoferritin metacrystal, one can easily approach the transition into the magnetically ordered phase by lowering the lattice temperature, and may hope to observe transient ordering when the superspin temperature, driven by the oscillating field, crosses the transition point. The theory should then also be modified to account for the onset of magnetic order [28].

The experimental method of detection the second harmonic of the high-frequency magnetic response as a function of static magnetic field [24] has proved very useful for detection of magnetic nanoparticles in biological tissues [26]. It can be modified for application to living organisms. Magnetoferritin metacrystals are considered prospective for application in magnetogenetics, i.e.

manipulation of biological processes (such as e.g. neural transduction) by externally applied magnetic field [6]. To this end, metacrystals should be injected into the organism and directed to a target receptor by biochemical and genetic manipulations. Their state while on target should be controlled in a non-perturbative way, for which purpose the nonlinear magnetic response seems very prospective.

Naturally formed magnetoferritin metacrystals were recently discovered in the inner ear of several bird species, while they are absent in mammals [10]. The spherical objects with the ferritin or magnetoferritin metacrystal structure, called cuticulosomes [11], have 0.5 micrometers in diameter and are situated in hair cells that provide sensitivity to vibrations in the hearing organ and to accelerations and gravity in the vestibular system [30]. Since birds are known to use magnetic compass for orientation during migration flights [31, 32], the possible role of these magnetic inclusions in magnetoreception has been discussed [10,12], though no plausible mechanisms were so far proposed. The magnetic susceptibility of the cuticulosome is apparently insufficient to provide sensitivity to the geomagnetic field [12]. Whether or not rearrangement of the ferritin balls filling the cuticulosome and, possibly, their transfer to mechanosensitive hairs, the stereocilia [30], could provide such a sensitivity, depends on the magnetic moment of the ferritin core, which is so far unknown. Measurement of the nonlinear high-frequency response and comparison of the results with the above theory may help to solve this problem, as well as to provide screening of other parts of body in birds and other organisms for similar structures.

*5. Conclusions.*

We have developed a theory of the high-frequency magnetic response of the metacrystal composed of magnetoferritin macromolecules containing superparamagnetic magnetite cores. The theory is applicable at high enough temperatures (in particular, at room temperature), when the metacrystal is in the paramagnetic phase. It is based on the observation that the time of establishing the internal equilibrium within the magnetic system of the metacrystal, which occurs due to the magnetic dipole-dipole interaction of magnetite cores, is more than an order of magnitude shorter than the time of its Néel relaxation due to lattice vibrations. This makes the magnetic system of the metacrystal similar to the nuclear spin system of a dielectric solid, for which the spin temperature approach is known to adequately describe the vast majority of experimental facts. The dissimilarity of the two systems is related to their relaxation timescales. For the magnetoferritin metacrystal it is measured in nanoseconds, which limits the range of relevant effects to the response of the magnetic system to rapid (with the frequencies lying in the megahertz range) variations of the strength of the applied magnetic field. For this reason, the basic equation of the theory is a differential equation for the inverse superspin temperature as a function of time. It is derived from the balance of energies under the condition that the magnetic system reaches its internal equilibrium fast enough to follow the external magnetic field variations. The equation allows one to compute the time-dependent magnetization for an arbitrary amplitude of the external field. In particular, it gives the complex amplitudes of higher harmonics of the applied sinusoidal probe field as functions of the background static magnetic field. These functions can be used for characterization of artificial magnetoferritin metacrystals, as well as for detection and non-destructive control of such objects in biological tissues. Application of this technique for studying the cuticulosomes, natural ferritine or magnetoferritine metacrystals of spherical shape, presents a special interest, since cutuculosomes are found in receptor cells in the inner ear of birds and might bear some relation to their ability to use geomagnetic field for orientation during seasonal migrations.

## Acknowledgements

The author is grateful to V.A.Ryzhov, M.M.Glazov and D.S.Smirnov for very helpful discussions. This work was supported by grant no. 16-14-10159 from Russian Science Foundation.